\documentclass[article]{elsarticle}

\usepackage{lineno,hyperref}

\journal{Journal of \LaTeX\ Templates}

\begin{document}

\begin{frontmatter}

\title{ISR corrections to associated $HZ$ production \\
at future Higgs factories}

\author[mainaddressRM,secondaryaddressRM]{Mario Greco}
\author[mainaddressPV,secondaryaddressPV]{Guido Montagna}
\author[secondaryaddressPV]{Oreste Nicrosini}
\author[secondaryaddressPV]{Fulvio Piccinini\corref{mycorrespondingauthor}}
\cortext[mycorrespondingauthor]{Corresponding author}
\ead{fulvio.piccinini@pv.infn.it}
\author[mainaddressPV]{Gabriele Volpi}

\address[mainaddressRM]{Dipartimento di Matematica e Fisica, Universit\`a di Roma 3, \\ 
via della Vasca Navale 84, 00146 Roma -- Italy}
\address[secondaryaddressRM]{INFN, Sezione di Roma 3, via della Vasca Navale 84, 00146 Roma -- Italy}
\address[mainaddressPV]{Dipartimento di Fisica, Universit\`a di Pavia, via A. Bassi 6, 27100 Pavia -- Italy}
\address[secondaryaddressPV]{INFN, Sezione di Pavia, via A. Bassi 6, 27100 Pavia -- Italy}

\begin{abstract}
We evaluate the QED corrections due to initial state radiation (ISR) to  
associated Higgs boson production in electron--positron ($e^+ e^-$) annihilation
at typical energies of interest for the measurement of the Higgs properties 
at future $e^+ e^-$ colliders, such as CEPC and FCC--ee.
 We apply the QED Structure Function approach  to the four--fermion 
 production process $e^+ e^- \to \mu^+ \mu^- b \bar b$, including both 
 signal and background contributions. We emphasize the relevance of the 
 ISR corrections particularly near threshold and show that finite third order 
 collinear contributions
 are mandatory to meet the expected experimental accuracy. 
 We analyze in turn the r\^ole played by a full four--fermion calculation 
 and beam energy spread in precision calculations for Higgs physics
 at future $e^+ e^-$ colliders.
\end{abstract}


\end{frontmatter}


\section{Introduction}

The discovery of a new scalar particle at the LHC in 2012 by the ATLAS~\cite{ATLAS2012} and CMS collaborations~\cite{CmS2012} has opened
 a new chapter in particle physics, and immediately has triggered the question of the real nature of this boson. 
 The determination of  the spin--parity quantum numbers and the couplings to other Standard Model (SM) particles strongly suggest it to be the 
 Higgs boson, i.e. the elementary scalar particle responsible for the mechanism of 
 electroweak symmetry breaking. However, from the available data it can not be concluded yet that we have found the SM Higgs boson 
 and not, for instance, one of the scalars postulated within possible extensions of the SM. 

Therefore, various lepton collider Higgs factories~\cite{Alexahin:2013-2, Aicheler:2012, BBFG:2013, Rubbia:2013, Bicer:2014, Ahmad:2015, Fujii:2017vwa, Asai:2017pwp} 
are under consideration to study in detail the properties of the new particle to great accuracies, because of the much more favorable experimental 
environment than that at hadron colliders. Amongst the many candidates of Higgs factories, one can distinguish two main categories. 

The first one exploits the possibility of $s$--channel Higgs resonant production, which  is especially important, due to the narrow width of the Higgs boson, 
of about 4 MeV, as predicted by the  SM. In particular, a muon collider Higgs factory~\cite{Alexahin:2013-2, Rubbia:2013}  
could produce the Higgs boson in the $s$--channel and perform an energy scan to map out the Higgs resonance line shape 
at tens of MeV level~\cite{BBGH:97}. 
This approach would provide the most direct measurement of the Higgs boson total width, the Yukawa coupling to muons and other fermions, 
and would also enable to simply investigate the existence of other scalar bosons predicted by natural extensions of the SM.  
This case has been studied in detail in Refs.~\cite{Greco:2015, Greco:2016izi, Jadach:2015cwa}, with a particular emphasis on the r\^ole of 
the Initial State Radiation (ISR) effects -- which are quite important~\cite{Greco:1975rm, Greco:1975ke} due to the $s$-channel resonant production 
of the very narrow Higgs boson -- and the evaluation of the background processes.

The second category includes the possibility of ultra--high luminosity electron--positron ($e^+ e^-$) colliders, 
such as the CEPC, FCC--ee,  the International Linear Collider and CLIC, which 
have been proposed~\cite{Aicheler:2012, BBFG:2013, Bicer:2014, Ahmad:2015, Fujii:2017vwa, Asai:2017pwp}
with the precise aim of observing the Higgs signal mainly through the reaction 
$e^+ e^- \to HZ$ at different center of mass (c.m.) energies, but also 
measuring possibly the Yukawa coupling to electrons.  In these cases, the ISR effects  could be quite sizable because
 of the smallness of the electron mass,  in particular in the vicinity of the threshold of $HZ$ production. A 
 study at LEP time of those effects, as well as of the background processes, was given in Ref.~\cite{MNP:1995}. 

In this paper, we focus on the process of associated $HZ$ production in $e^+ e^-$ annihilation, with the aim of providing 
a comprehensive and accurate theoretical approach to precision measurements of the Higgs 
boson parameters at future Higgs factories.
To this end, we consider for definitiveness the cleanest production channel given by the four--fermion 
process $e^+ e^- \to \mu^+ \mu^- b \bar b$, including both 
 signal and background contributions, and induced in the 
 signal process by the dominant Higgs decay $H \to b \bar{b}$ and the 
 leptonic $Z$ boson decay $Z \to \mu^+ \mu^-$. To model photon emission in the ISR process, we 
 use a set of representative and state--of--the--art  choices for the QED Structure 
 Functions, as largely employed in the context of
LEP precision phenomenology.

Next--to--leading electroweak corrections to $e^+ e^- \to HZ$ were calculated time ago 
in Refs.~\cite{Fleischer:1982af, Kniehl:1991hk, Denner:1992bc, Belanger:2002ik}, while 
mixed QCD--electroweak contributions have been computed recently in Refs.~\cite{Gong:2016jys, Sun:2016bel}. 
A very preliminary investigation of the ISR contribution to the $HZ$ signal process
at $\sqrt{s} = 250$~GeV has been performed in Ref.~\cite{Chen:2017ipx} using \textsc{MadGraph}.
Hence, our study improves the existing analyses of associated Higgs boson production at the proposed 
$e^+ e^-$ Higgs factories. 
The case of resonant Higgs production in electron-positron collisions and the modifications induced by 
ISR corrections on the Higgs boson line shape have been recently studied in Refs.~\cite{Greco:2016izi, Jadach:2015cwa}.

The paper is organized as follows. In Sect.~2, we describe the formulation and parameterization of the ISR 
corrections in the general framework 
of the QED Structure Function approach. In Sect.~3, we quantify their effects on the Higgs boson associated 
production cross section at various 
c.m. energies of interest for the different projects of Higgs factories.  We analyze in turn the r\^ole played by a full four--fermion calculation 
 and beam energy spread in precision calculations for Higgs physics
 at future lepton colliders.~\footnote{We do not consider in our study the 
 contribution of beamstrahlung, its effect being largely dependent from the considered 
 Higgs factory project.}
Our conclusions and perspectives are drawn in Sect.~4.

\section{Structure Function formulation of ISR effects}

According to factorization theorems of soft and collinear singularities, 
the contribution of ISR to Higgs boson associated production can be evaluated using the following master formula: 
\begin{equation}
{\rm d} \sigma (s) = \int {\rm d} x_1 {\rm d} x_2 D(x_1, s) D(x_2, s) {\rm d} \sigma_0 (x_1 x_2 s) \Theta ({\rm cuts}) . 
\label{eq:master}
\end{equation}
In Eq.~(\ref{eq:master}), ${\rm d} \sigma_0 $ is any  tree--level differential 
cross section including the signal $e^+ e^- \to Z H \to \mu^+ \mu^- b \bar b$ and all the background matrix elements, 
as computed in Ref.~\cite{MNP:1995}~\footnote{The complete four--fermion 
calculation of the background contributions includes 24 Feynman diagrams corresponding to the neutral--current 
processes of $\gamma \gamma, \gamma Z$ and $Z Z$ production and decay~\cite{MNP:1995}. 
We compute them and the signal matrix element in the fermion massless approximation.}, taken at the reduced squared c.m. energy $x_1 x_2 s$. 
The $ \Theta$ function represents the imposed kinematical cuts. The function $D(x, s)$ is the 
non--singlet  collinear Structure Function modeling initial state photon radiation and 
giving the probability of finding inside a parent electron an electron with momentum
fraction $x$ at the energy scale $s$. They were first introduced in Ref.~\cite{Kuraev:1985} and later improved for 
precision physics at LEP in Ref.~\cite{Nicrosini:1987} to add second order finite contributions
 to the resummation of leading logarithmic corrections. More recently, also third order finite terms have been 
 computed analytically in Refs.~\cite{Cacciari:1992, Skrzypek:1991, Skrzypek:1992, Montagna:1997}. 
 In particular, in Ref.~\cite{Cacciari:1992} the explicit 
 analytical expression of finite additive third order terms was given, together with an iterative 
 formula to compute higher and higher order contributions. The explicit analytical expression of finite 
factorized contributions can be found in Refs.~\cite{Skrzypek:1991, Skrzypek:1992}. In Ref.~\cite{Montagna:1997} the analytical expression for the
 radiator function, defined as a convolution 
of two Structure Functions, is given up to third order by using the Structure Functions of Ref.~\cite{Cacciari:1992}. 
In Ref.~\cite{Arbuzov:1999} finite 
forth and fifth order additive finite terms are provided in distributional form. A review of the QED Structure Function method can be found 
in Refs.~\cite{Montagna:1998, Arbuzov:2010zzb}. 

The explicit expressions of the Structure Functions used in the present study are listed in the following. The all--order 
Structure Function, valid in the soft photon limit and dubbed as Gribov--Lipatov solution, is given by
\begin{equation}
D_{GL} (x,s ) = \frac{\exp\left[  \frac{1}{2} \beta \left( \frac{3}{4} - \gamma_E \right) \right]}{\Gamma \left( 1 + \frac{1}{2} \beta \right) }
\frac{1}{2} \beta \left( 1 - x \right)^{\frac{1}{2} \beta - 1} , 
\label{eq:DGL}
\end{equation}
where 
\begin{equation}
\beta =  \frac{2 \alpha}{\pi} (L - 1), \qquad L = \ln(s/m_e^2) .
\end{equation}
In the above equations, $\alpha$ is the fine structure constant, $m_e$ is the electron mass, $\Gamma$ and $\gamma_E$ are the Gamma function 
and the Euler--Mascheroni constant, respectively. 
According to Eq.~(\ref{eq:DGL}) and following equations, photon radiation is treated in strictly collinear approximation and 
the collinear logarithmic enhancements are encoded in the large $\beta$ factor, with 
$\beta \simeq 0.11$ at the $HZ$ threshold.
The additive Structure Function including up to third order finite terms reads as follows~\cite{Cacciari:1992}: 
\begin{eqnarray}
D_A (x,s) &=& \sum_{i=0}^3 d_A^{(i)} (x,s) , \nonumber \\
d_A^{(0)} (x,s) &=& D_{GL} (x,s) , \nonumber \\
d_A^{(1)} (x,s) &=& - \frac{1}{4} \beta (1+x) , \nonumber \\
d_A^{(2)} (x,s) &=& \frac{1}{32} \beta^2 \left[ \left(1+x \right) \left(-4 \ln (1-x) + 3 \ln(x) \right) 
-4 \frac{\ln x}{1-x} -5 -x \right] , \nonumber \\
d_A^{(3)} (x,s) &=& \frac{1}{384} \beta^3 \Bigg\{ (1+x) \left[ 18 \zeta(2) -6 {\rm Li}_2 (x) -12 \ln^2 (1-x) \right]  \nonumber \\
&+& \frac{1}{1-x} \left[ - \frac{3}{2} (1 + 8 x + 3 x^2) \ln x -6 (x+5) (1-x) \ln(1-x) \right.  \nonumber \\
&-& 12 (1+x^2) \ln x \ln(1-x) + \frac{1}{2} (1+7 x^2) \ln^2 x \nonumber \\
&-& \left.  \frac{1}{4} (39 - 24 x - 15 x^2) \right] \Bigg\} . 
\label{eq:Dadd}
\end{eqnarray}
where $\zeta$ is the Riemann $\zeta$ function and ${\rm Li}_2$ is the dilogarithm.
The factorized Structure Function with up to third order finite terms can be written as~\cite{Skrzypek:1991, Skrzypek:1992}: 
\begin{eqnarray}
D_F (x,s) &=& D_{GL} (x,s) \sum_{i=1}^3 d_F^{(i)} \nonumber \\
d_F^{(1)} (x,s) &=& \frac{1}{2} (1 + x^2) , \nonumber \\
d_F^{(2)} (x,s) &=& \frac{1}{4} \frac{\beta}{2} \left[ - \frac{1}{2} (1 + 3 x^2) \ln x - (1-x)^2 \right] , \nonumber \\
d_F^{(3)} (x,s) &=& \frac{1}{8} \left( \frac{\beta}{2} \right)^2 \left[ (1-x)^2 + \frac{1}{2} (3x^2 - 4x+1) \ln x \right. \nonumber \\
&+& \left. \frac{1}{12} (1+7x^2) \ln^2 x + (1-x^2) {\rm Li}_2 (1-x) \right]
\label{eq:Dfac}
\end{eqnarray}
In the next Section, a detailed comparison between the results obtained by using all the above Structure Functions 
for the modeling of ISR corrections to associated Higgs production is discussed. 

\section{Numerical results}

For the presentation of the numerical results, we consider two distinct event selection conditions:
\begin{enumerate}

\item No Cuts. In this situation, we simply require that the invariant masses of the $\mu^+ \mu^-$ and $b \bar{b}$ pairs 
satisfy the loose constraint $M_{\mu^+ \mu^-} \geq 12$~GeV and $M_{b \bar{b}} \geq 12$~GeV. These cuts are 
imposed in order to suppress the contribution from the background processes of $\gamma\gamma$ and $Z\gamma$ 
production and decay.

\item Cuts. According to this event selection, in order to mimic the finite $b \bar{b}$ mass resolution foreseen at 
future $e^+ e^-$ colliders, we apply a cut on the $b \bar b$ pair invariant mass 
given by $  M_H - 3~{\rm GeV} \leq m_{b \bar b} \leq M_H + 3~{\rm GeV}$~\cite{Tenchini:2017}, with $M_H$ = 125~GeV, in 
association to the cut $M_{\mu^+ \mu^-} \geq 12$~GeV.

\end{enumerate}

We use as input parameters $M_Z = 91.1876$~GeV, $\Gamma_Z = 2.4952$~GeV, $M_W = 80.385$~GeV, $M_H = 125$~GeV 
and $m_b = 4.7$~GeV, the latter being necessary for the evaluation of the $Hb\bar{b}$ Yukawa coupling. All the other derived quantities 
are computed at the three level, using the so--called $G_\mu$ scheme.

\begin{figure}[hbtp]
\centering
\includegraphics[width=11.cm]{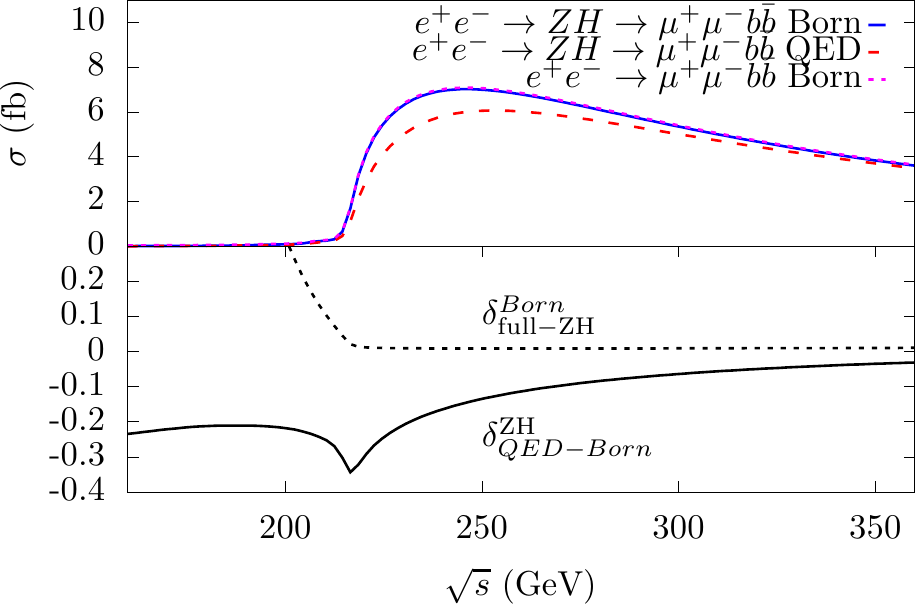}
\caption{The total cross section of the process $e^+ e^- \to \mu^+ \mu^- b \bar b$ 
for the $ZH$ signal contribution in the Born approximation (Born), for the 
full four--fermion calculation in the Born approximation ($e^+ e^- \to \mu^+ \mu^- b \bar b$ Born) and with ISR QED corrections 
to the signal process (QED), as a function of the c.m. energy (upper panel). The relative effect of ISR corrections is shown in lower panel. 
Numerical results correspond to the event selection 2. described in the text.}
\label{Fig:1}
\end{figure}

From the numerical simulation, we observe that at $\sqrt{s} = 240$~GeV, in the absence of a cut 
on the $b \bar{b}$ invariant mass around the Higgs mass, the signal+background (S+B)
$\mu^+ \mu^-$ invariant mass distribution is about a factor of four larger than the signal alone (S), whereas the cut reduces the 
background such that the signal alone 
and the signal plus background $\mu^+ \mu^-$ invariant mass distributions 
differ at the per cent level. The corresponding cross sections at $\sqrt{s} = 240$~GeV are 
$\sigma_S = 5.899(3)$~fb and $\sigma_{S+B} = 24.39(2)$~fb when no cuts are imposed, while we obtain
$\sigma_S = 5.899(3)$~fb and $\sigma_{S+B} = 5.960(3)$ in the presence of a $b \bar{b}$ invariant mass cut 
around the Higgs mass. The digits in parenthesis correspond to the 1$\sigma$ Monte Carlo error estimate.
These cross section values are obtained by including ISR QED corrections in the 
Gribov--Lipatov approximation as in Eq.~(\ref{eq:DGL}). This simple analysis points out that 
the theoretical predictions must rely upon a full 
four--fermion calculations in order to meet the foreseen experimental precision, which
is presently estimated at the level of $0.4 \times 10^{-3}$~\cite{Piccinini:2017}, at least 
for the CEPC and FCC--ee facilites.

The relevance of ISR QED contributions to associated Higgs boson production, neglected in previous studies, is shown in 
Fig.~\ref{Fig:1}, where the line shape of the signal process $e^+ e^- \to Z H \to \mu^+ \mu^- b \bar b$ in the lowest order 
approximation (Born) is compared to the corresponding QED corrected cross section (QED) 
computed according to Eq.~(\ref{eq:master}) 
(upper panel); the relative impact of ISR corrections is shown in the lower panel of Fig.~\ref{Fig:1}. The contribution
of ISR has been evaluated by using the Structure Function of Eq.~(\ref{eq:Dadd}) including  finite third order contributions. 
We also show in Fig.~\ref{Fig:1} the line shape of the full four--fermion calculation in the tree--level 
approximation ($e^+ e^- \to \mu^+ \mu^- b \bar b$ Born). These results have been obtained according to the event selection condition 2. above.
It can be noticed that the overall effect of ISR QED corrections is 
strongly varying with $\sqrt{s}$ and quite large, at the level of -35\% in the threshold region and 
between -10\% and -20\% at several c.m. energy values. The large impact of the ISR corrections at threshold can be 
simply understood in terms of the cut--off on the maximum photon energy induced by the finite width 
of the $Z$ and $H$ bosons and therefore mainly derives from the contribution of multiple soft photon radiation. 
Incidentally, one can also see, as already remarked, that the full four--fermion prediction differs from the signal calculation 
at the per cent level at and above threshold, thus emphasizing the importance of a full $4f$ 
calculation also in the presence of a tight cut on the invariant mass of the $b \bar{b}$ fermion pair.

 \begin{figure}[hbtp]
\centering
\includegraphics[width=11.cm]{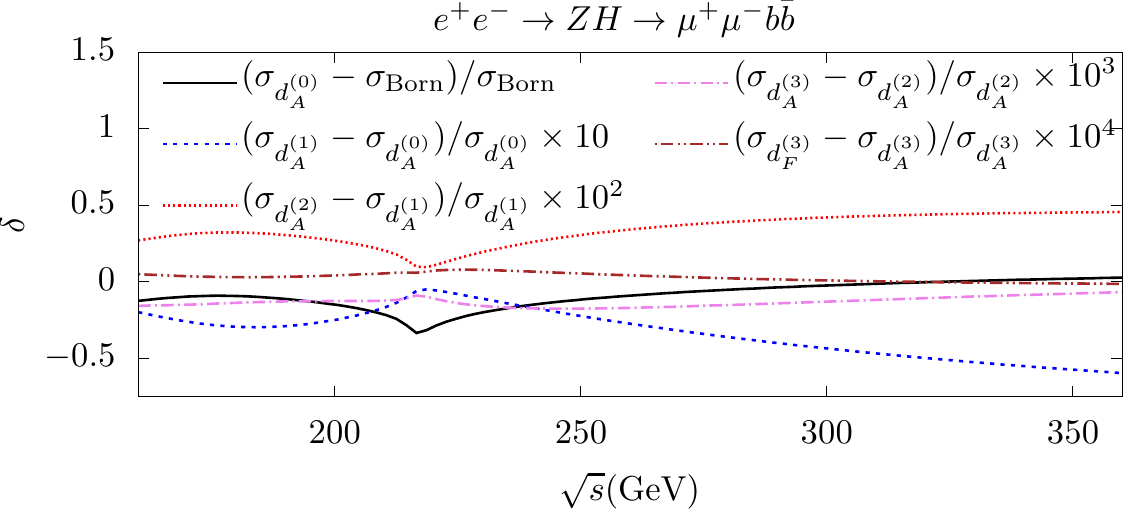}
\caption{The relative effect of different approximations for the electron Structure Functions modeling ISR corrections 
to the $e^+ e^- \to Z H \to \mu^+ \mu^- b \bar{b}$ signal process, as a function of 
the c.m. energy. Results corresponding to the event selection 2. described in the text.}
\label{Fig:2}
\end{figure}

Figure~\ref{Fig:2} shows the relative effect of the different finite order approximations 
included in the electron Structure Function. The main conclusions that can be drawn from Fig.~\ref{Fig:2}
are the following. 
The first order additive finite effects, as compared to the 
Gribov--Lipatov approximation describing multiple soft photon emission, 
are of the order of some percent, increasing with the c.m. energy above threshold. 
This can be understood as hard photon radiation becomes more and more important
 well above threshold. The second order additive finite contributions, as compared to the up to first order ones, 
 amount to some per mille and are therefore strictly necessary for a precision measurement
 of the Higgs production cross section. The third order additive finite effects are at the level of $10^{-4}$ in comparison with the Structure Function containing 
up to second order contributions. In view of the foreseen precision at future Higgs factories, these $O(\beta^3)$ 
collinear effects have to be carefully taken into account. 
A final remark concerns the relationship between the additive corrections of Eq.~(\ref{eq:Dadd}) and the 
factorized contributions of Eq.~(\ref{eq:Dfac}). The comparison 
between the two prescriptions,  as illustrated in Fig.~\ref{Fig:2}, 
shows that the relative difference between the third order additive and factorized corrections is 
at the level of $10^{-5}$ or below it, and thus negligible
 as compared to the foreseen experimental precision. In other words, the two approximations can be considered as equivalent in 
comparison with the expected experimental accuracy.
Since the additive and factorized Structure Functions including finite effects at a given order 
$\beta^n$ differ for finite terms at the next order $\beta^{n+1}$, one can take the difference between third order 
additive and factorized solutions as an estimate of fourth order finite contributions. Should in a future a higher precision 
in the calculation of ISR QED corrections be needed, higher order finite effects could be included using the available
 analytical expressions for the fourth and/or fifth order collinear contributions given in Ref.~\cite{Arbuzov:1999}.  
 
\begin{figure}[t]
\centering
\includegraphics[width=11cm]{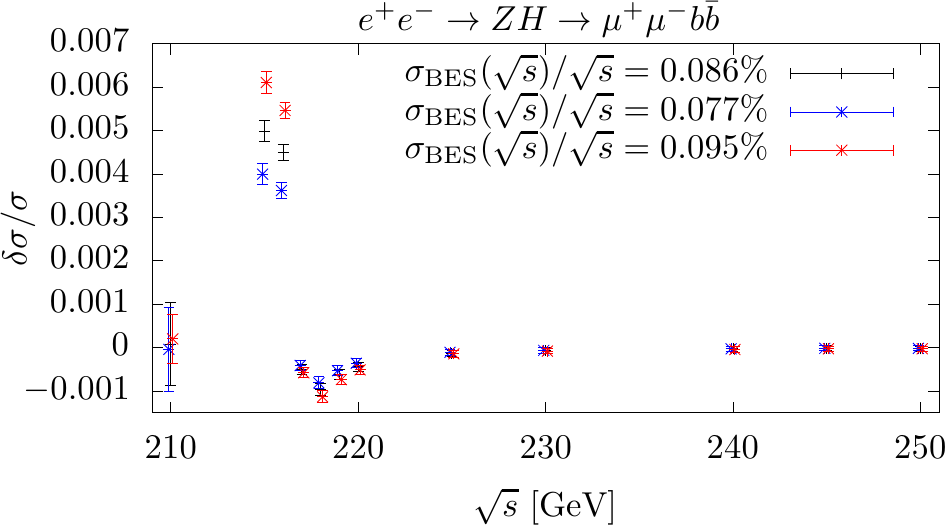}
\caption{Contribution of the beam energy spread and related uncertainty to the cross section of the signal process 
$e^+ e^- \to Z H \to \mu^+ \mu^- b \bar{b}$, as a function of the c.m. energy.}
\label{Fig:3}
\end{figure}

In Fig.~\ref{Fig:3} we show, for completeness, the impact due to the beam energy spread and related uncertainty 
on the cross section of the $HZ$ signal process as a function of $\sqrt{s}$. We simulated the profile
of the machine energy spread in terms of a Gaussian distribution of the c.m. energy, with a standard deviation 
$\sigma_{\rm BES}$ associated to the energy spread given by $\sigma_{\rm BES} (\sqrt{s}) / \sqrt{s} = (0.086 \pm 0.009)$\%, where the 
uncertainty corresponds to a knowledge of about 10\% to its value~\cite{Blondel:2017}. As it can be seen 
from Fig.~\ref{Fig:3},  the contribution of the machine energy spread is only relevant at threshold, being of the order 
of 0.5\% with an uncertainty of about 0.05\%, whereas it is negligible elsewhere, below the $10^{-4}$ level.

\section{Conclusion}

We have computed the  ISR QED corrections to the process $e^+e^- \to \mu^+ \mu^- b \bar b$ 
from $HZ$ associated production at typical energies of interest for the measurement of the properties of the Higgs particle 
at future $e^+ e^-$ facilities. Using different prescriptions for the electron Structure Function at various 
levels of sophistication, 
we have shown that the QED radiative corrections are quite substantial, especially in the vicinity of the $HZ$ threshold. 
We have also provided clear evidence that third order collinear contributions must be taken into account in order to meet the 
expected experimental accuracy of future Higgs factories. We have also evaluated the impact due to the 
beam energy spread 
and related uncertainty, to conclude that this effect is only relevant at threshold but is not in general a limiting factor in 
precise predictions  for associated Higgs boson production at future $e^+ e^-$ accelerators.

Our study improves the existing analyses of the proposed Higgs factories and can serve as a guideline for the 
target accelerator designs with respect to the physics goals.

Possible perspectives of the present work include the evaluation of ISR corrections to other relevant signatures 
for physics at future $e^+ e^-$ accelerators, such as $WW$, $t \bar{t}$, $ZHH$ and 
$t \bar{t} H$ production processes.

\vskip 12pt \noindent
{\bf Acknowledgements}
\vskip 8pt\noindent
We are grateful to A. Blondel, P. Janot and R. Tenchini for useful discussions.

\bibliography{Higgsbibfile}

\end{document}